%% file: 0_IDC26.tex
\documentclass[sigconf]{acmart}
\AtBeginDocument{%
  }

\copyrightyear{2026}
\acmYear{2026}
\setcopyright{cc}
\setcctype{by-nc-nd}
\acmConference[IDC '26]{Proceedings of the 25th Interaction Design and Children Conference}{June 22--25, 2026}{Brighton, United Kingdom}
\acmBooktitle{Proceedings of the 25th Interaction Design and Children Conference (IDC '26), June 22--25, 2026, Brighton, United Kingdom}
\acmDOI{10.1145/3773077.3812156}
\acmISBN{979-8-4007-2283-7/2026/06}

\usepackage{booktabs}
\usepackage{tabularx}
\usepackage{multirow}
\usepackage{amsmath}

\begin{document}

\title[Understanding teens’ self-beliefs when learning to construct and deconstruct
AI/ML systems]{Understanding teens’ self-beliefs when learning to construct
and deconstruct AI/ML systems: Developing a survey instrument}


\author{Luis Morales-Navarro}
\orcid{0000-0002-8777-2374}
\email{luismn@upenn.edu}
\affiliation{%
  \institution{University of Pennsylvania}
  \city{Philadelphia}
  \state{Pennsylvania}
  \country{USA}
}

\author{Deborah Fields}
\orcid{0000-0003-1627-9512}
\email{deborah.fields@usu.edu}
\affiliation{%
  \institution{Utah State University}
  \city{Logan}
  \state{Utah}
  \country{USA}
}

\author{Michael T. Giang}
\orcid{0009-0000-9050-5704}
\email{mtgiang@cpp.edu}
\affiliation{%
  \institution{Cal Poly Pomona}
  \city{Pomona}
  \state{California}
  \country{USA}
}

\author{Daniel J. Noh}
\orcid{0009-0002-7219-1988}
\email{dnoh@upenn.edu}
\affiliation{%
  \institution{University of Pennsylvania}
  \city{Philadelphia}
  \state{Pennsylvania}
  \country{USA}
}

\author{Yasmin B. Kafai}
\orcid{0000-0003-4018-0491}
\email{kafai@upenn.edu}
\affiliation{%
  \institution{University of Pennsylvania}
  \city{Philadelphia}
  \state{Pennsylvania}
  \country{USA}
}

\author{Dana{\'e} Metaxa}
\orcid{0000-0001-9359-6090}
\email{metaxa@seas.upenn.edu}
\affiliation{%
  \institution{University of Pennsylvania}
  \city{Philadelphia}
  \state{Pennsylvania}
  \country{USA}
}

\renewcommand{\shortauthors}{Morales-Navarro et al.}

\begin{abstract}
  \input{1_sections/0-abstract}
\end{abstract}


\maketitle

\input{1_sections/1-intro}
\input{1_sections/2-background}
\input{1_sections/3-methods}
\input{1_sections/4-findings}
\input{1_sections/5-discussion}
\input{1_sections/6-conclusion}

\section{Selection and Participation of Children}
We recruited teens enrolled in computing courses in the Western United States. Teens were invited by their teachers to participate in the study. Parents received consent forms prior to the study, which included a brief explanation of the research, and a researcher visited the classrooms to answer any questions teens had about the study. With IRB and district approval, surveys waived consent/assent as they collected no identifying information. Research protocols and data collection methods were approved by the IRB board of the University of Pennsylvania.

\begin{acks}
This work was partially supported by National Science Foundation grants (\#2414590, \#2333469), a Penn AI fellowship, and a rapid response grant from the Spencer Foundation, the Kapor Center, the William T. Grant Foundation, and the Alfred P. Sloan Foundation.

\end{acks}

\bibliographystyle{ACM-Reference-Format}
\bibliography{4_references}
\end{document}

%% file: 1_sections/0-abstract.tex
Despite growing calls to foster AI literacy, there are few available survey instruments designed for children and youth that study computational empowerment alongside construction and deconstruction activities. In such activities, learners' beliefs about their abilities and attributes can impact their engagement. In this paper, we introduce and validate a survey instrument with constructs related to construction (creative expression and problem-solving self-beliefs) and deconstruction (auditing self-efficacy and fascination with auditing), along with more general self-beliefs related to design justice and the value of learning about AI/ML. We administered the instrument to 124 teenagers and assessed the six-factor structure of the instrument using confirmatory factor analysis. In addition to confirming the structure, we found that design justice beliefs strongly correlated with problem-solving, auditing self-efficacy, and creative expression.


%% file: 1_sections/1-intro.tex
\section{Introduction}

Efforts to introduce AI/ML literacies\footnote{We use the word \textit{literacies} in plural, recognizing the multiple literacies that have been proposed in relation to AI/ML from effective and responsible use to learning how AI/ML systems are designed and learning how to systematically evaluate them.} activities are growing in K-12 education around the globe \cite{oecd2025empowering}. Most of these efforts focus on helping children and youth to become more efficient users or improve their technical knowledge or understanding. Some of these efforts are framed from the perspective of computational empowerment, which supports young people in construction (design of applications) and deconstruction (evaluation of applications) activities \cite{smith2023research}. In such activities, learners' beliefs about their abilities and attributes can impact their engagement. At the same time, assessment of AI literacies interventions remains underdeveloped \cite{casal2023ai}. Most instruments focus on kids' technical knowledge or their understanding of AI/ML from the perspective of using the systems (rather than designing or evaluating systems) \cite{zhang2025developing, ng2024design}. We need to measure not only what learners understand but also their beliefs and dispositions towards learning to construct and deconstruct AI/ML systems.

In this paper, we introduce and validate a short survey with constructs related to construction (creative expression and problem-solving self-beliefs), deconstruction (auditing self-efficacy and fascination with auditing), and more general self-beliefs related to design justice and the value of learning about AI/ML. We administered the instrument to 124 teenagers and assessed the six-factor structure of the instrument using confirmatory factor analysis. In the discussion, we address how the different constructs were related to each other and propose future directions for the instrument development, including adaptations for younger learners.

%% file: 1_sections/2-background.tex
\section{Background and Related Work}
In child-computer interaction, many efforts to build AI literacies are framed from the perspective of computational empowerment (CE) \cite{smith2023research}. CE emphasizes that learning about computing should go beyond system use and technical understanding to provide learners with opportunities to participate in the development and evaluation of computing systems \cite{dindler2020computational, iversen2018computational}.

In the context of AI/ML, researchers have investigated cognitive aspects of how young people engage in the construction or design of AI/ML models and applications \cite{kahila2026engaging, morales2025high}. There is some evidence that such construction activities can support learners to better understand the AI/ML systems they use in their everyday lives \cite{druga2021children}. Researchers have also investigated how learners can deconstruct AI/ML systems by evaluating their performance through auditing activities \cite{solyst2025investigating} and reflecting on their limitations and implications \cite{bilstrup2023embodied}. This work shows that teens are able to lead audits from beginning to end \cite{morales2025learning} and understand some of the limitations and implications of AI/ML systems \cite{salac2023funds}. 

However, there is a need to better understand both cognitive and non-cognitive aspects of constructing and deconstructing AI/ML systems. In this paper, we center on teens' self-beliefs, consisting of an array of non-cognitive self-terms \cite{wylie1974self} (e.g., self-concept and self-efficacy) that emphasize the beliefs individuals hold about their own abilities and attributes \cite{valentine2004relation}, which can, in turn, influence engagement and performance in learning \cite{valentine2004relation}. Such self-beliefs are domain-specific. For example, self-beliefs about language arts do not necessarily apply to self-beliefs about computing \cite{bandura1997self, harter1999construction}. Within computing education research, self-beliefs related to efficacy, anxiety, enjoyment, confidence, belonging, and persistence have been investigated in the context of learning programming \cite{decker2019topical, lishinski201928}. However, as \citet{tedre2021ct} argue, learning about AI/ML is different from traditional programming, and as such, instruments must be adapted to account for model design and evaluation learning activities.

\subsection{Existing Instruments}

While many AI/ML literacies instruments have been developed, a review of different surveys shows that very few instruments are designed for teenagers, and the majority center on examining people’s beliefs about \textit{using} AI/ML systems, not designing and evaluating them \cite{lintner2024systematic}. Instruments for teens have focused on their conceptual understanding of AI/ML \cite{zhang2025developing}, self-efficacy, motivation, and confidence to use AI/ML systems \cite{ng2024design}. More recently, \citet{10.1145/3702652.3744208} developed an instrument for computing professionals and higher education students (average age 30) with scales about the value of marginalized perspectives, techno-solutionism, ethics training, technical training, personal effectiveness, and system responsiveness. Of these constructs, only one (personal effectiveness) centers on self-beliefs, asking responders to rate their agreement with statements such as ``I am able to participate in discussions focusing on ethics and social impacts of computing.'' Further research is needed to develop similar constructs that can be used with teens to measure their self-beliefs when constructing and deconstructing AI/ML systems. 

In this paper, we present and validate an instrument with constructs adapted and adopted from prior research in computing education that centers on self-beliefs \cite{scott2014measuring, morales2024connecting, lishinski201928}. In particular, we draw heavily from constructs in the CS Interests and Beliefs Inventory (CSIBI), an instrument previously developed by our team \cite{morales2024connecting}. These changes include redesigning items to focus on AI/ML rather than programming, as well as adding constructs that focus on deconstruction, particularly AI auditing. We also include a new construct to measure design justice self-beliefs. In section \ref{sec:instrument}, we describe each of the constructs of the instrument. 

%% file: 1_sections/3-methods.tex
\section{Methods}

\subsection{Context and Participants}
This study was conducted within a larger participatory design project on AI auditing, grounded in principles that foreground teacher and youth voices in computing education \cite{smith2023research}. During the year prior to classroom data collection, five computer science teachers collaborated with researchers on developing a set of lessons for high school (grades 9-12) CS classrooms to foster critical engagement with AI/ML systems through AI auditing \cite{noh2025advisor, noh2026teacher}. In Spring 2025, three of the teachers at three different schools (in two school districts) located on the West Coast of the United States piloted the first draft of these lessons over 2-3 weeks ($\sim$10-15 hours of class time). Two teachers piloted the lessons in introductory computing classes (Exploring Computer Science, see \cite{goode2012ecs}), primarily with 9th or 10th graders. One teacher piloted the lessons in the most advanced computing class in his school, an Introduction to Data Science for 11th and 12th graders. 

Each of the three public schools served primarily under-resourced youth (56-92\% free and reduced lunch as a measure of socioeconomic status) in ethnically diverse, metropolitan communities (primarily Latinx, Black, White, Polynesian, Asian, and Multiracial; 70-99\% non-White). All three teachers had 9-12 years experience teaching computer science (15-22 years teaching experience overall) and were solicited for the study in part because of their extensive years of experience building equity in CS within their specific school communities.

\subsection{The Instrument}
\label{sec:instrument}
The instrument we piloted has six different constructs. In the following paragraphs, we explain how these constructs were created, adapted, and adopted. 

\textbf{Fascination with Auditing} captures positive affect, interest, and curiosity towards auditing. Fascination relates to learners' emotional and cognitive attachment to specific topics, which can encourage participation and drive career interests \cite{dorph2018grow}. Items were adapted from CSIBI \cite{morales2024connecting}, which used the Activation Lab STEM Fascination scale.  ``I talk about the process of auditing machine learning systems with friends or family'' is an example item.

\textbf{Value of Learning AI/ML} measures learners’ beliefs about the value of learning about AI/ML. We adapted selected items from the CSIBI, which in turn were adapted from the Activation Lab STEM Values scale \cite{chung2016valuing} (e.g., ``Learning computing and machine learning is important for me in the future.'').

\textbf{Computing Creative Expression} is not AI/ML specific and was adopted from CSIBI to measure learner beliefs about being able to express themselves creatively with computing (e.g., ``I can make things that are interesting to me in computing.''). Such a construct is of importance among teens since designing personally relevant artifacts is widely used to promote engagement with computing \cite{oleson2020role, harel1990software}.

\textbf{Auditing Self-efficacy} is an adaptation of the coding self-efficacy construct in CSIBI (e.g., from ``I am confident that I can understand Arduino errors'' to ``I am confident that I can systematically evaluate the outcomes of a machine learning system, like the one I worked with in this session/workshop.''), which in turn is based on Scott and Ghinea's CS1 self-efficacy \cite{scott2014measuring}.

\textbf{Design Justice} is a new construct based on \citeauthor{costanza2020design}'s design justice. 
Design justice is concerned ``with how the design of objects and systems influences the distribution of risks, harms, and benefits among various groups of people'' \cite{costanza2018design}. It measures teens' justice-oriented self-beliefs in relation to design processes. An example item is ``When finding and fixing problems in my projects, I think about how these issues affect people.''

\textbf{Problem-solving} was adopted from CSIBI, addressing learners’ confidence in their ability
to solve problems in computing. The items are based on the Activation Lab STEM Competency Beliefs scale \cite{chen2017measures} and were adapted to computing (e.g., from ``I think I am very good at explaining my solutions to math problems'' to ``I think I am very good at explaining my solutions to technical problems'').

\subsection{Data Collection}
At the end of the pilot lessons, students filled out the survey instrument either digitally or on paper. Digital surveys (through Qualtrics) presented items and constructs in randomized sequences. Paper surveys (one district only allowed paper surveys) consisted of five versions with items and constructs in different sequences, handed out in random order to students. The number of classes in which teachers implemented the lessons depended on their course schedules, with one teacher implementing the lessons in five introductory CS courses, and two teachers in one course each. In total, 124 secondary school students completed the survey across 7 classes (N=124; 87 digital, 37 paper). Individual student demographics were not collected by this instrument. We manually entered paper survey responses into Qualtrics for analysis.

\subsection{Analysis}

To evaluate the six-factor structure for the 21-item instrument, a maximum likelihood confirmatory factor analysis (CFA) was performed using \texttt{lavaan} in R \cite{rosseel2012lavaan}. The CFA analysis tests whether the number of proposed factors fits the data, each indicator item adequately loads onto the proposed factor, the errors (or uniqueness) across each indicator are unrelated, and the relationship (correlations) exists across all the latent factors. As there are no definitive model fit criteria for CFA, multiple indicators were used. For this study, adequate model fit indicators were chi-square goodness of fit (p > .05 \text{ or } x/df < 3.00), root mean square error of approximation 2 (RMSEA < .06), standardized root mean square residual (SRMR < .08), comparative fit index (CFI > .90), and Tucker–Lewis index (TLI > .90).

%% file: 1_sections/4-findings.tex
\section{Findings}

Confirmatory factor analysis results indicated that the six-factor model had an acceptable fit (CFI = 0.91, TLI = 0.891, RMSEA = 0.076, SRMR = 0.062) for the data across model indices (except the chi-square p-value = 0.000, which is often significant with larger sample sizes). In addition, the factor loadings for each item averaged 0.755 (ranging between 0.614 and 0.923). All loadings were significant (p < 0.001). See Table \ref{tab:factor_loadings} for factor loadings and reliabilities for all constructs.

\begin{table*}[ht]
\centering
\caption{Factor Loadings, Factor Determinacy, and Reliability (Cronbach's $\alpha$)}
\label{tab:factor_loadings}
\small
\begin{tabularx}{\textwidth}{>{\raggedright\arraybackslash}X c c c c}
\toprule
\textbf{Factor / Item} & \textbf{Loading} & \textbf{S.E.} & \textbf{Det.} & \textbf{$\alpha$} \\
\midrule
\textbf{F1 - Fascination with Auditing} & & & 0.939 & 0.837 \\
I talk about the process of auditing ML systems with friends or family. & 0.923 & 0.079 & & \\
I talk about algorithmic justice with friends or family. & 0.781 & 0.081 & & \\
\addlinespace

\textbf{F2 - Value of Learning AI/ML} & & & 0.942 & 0.856 \\
Learning computing and ML is important for contributing to my community. & 0.734 & 0.063 & & \\
Learning computing and ML is important for me in the future. & 0.816 & 0.058 & & \\
Learning computing and ML is important for me in my future career. & 0.766 & 0.065 & & \\
I want to learn as much as possible about computing and ML. & 0.788 & 0.060 & & \\
\addlinespace

\textbf{F3 - Computing \& Creative Expression} & & & 0.935 & 0.852 \\
I can be creative in computing. & 0.823 & 0.059 & & \\
I can express myself in computing. & 0.808 & 0.062 & & \\
I can make things that are interesting to me in computing. & 0.803 & 0.063 & & \\
\addlinespace

\textbf{F4 - Auditing Self-efficacy} & & & 0.944 & 0.836 \\
I am confident that I can systematically evaluate ML system outcomes. & 0.698 & 0.063 & & \\
I am confident I can develop strategies to evaluate ML systems. & 0.757 & 0.055 & & \\
I am confident I can track inputs and outputs in ML systems. & 0.807 & 0.059 & & \\
I am confident I can identify potential biases in ML systems. & 0.741 & 0.059 & & \\
\addlinespace

\textbf{F5 - Design Justice} & & & 0.930 & 0.772 \\
Learning computing and ML can help me work with my community. & 0.788 & 0.049 & & \\
It is important to think about how what I create will be used by others. & 0.725 & 0.051 & & \\
When fixing problems, I think about how these issues affect people. & 0.706 & 0.049 & & \\
\addlinespace

\textbf{F6 - Problem-solving} & & & 0.919 & 0.810 \\
I am very good at figuring out how to fix things that don't work. & 0.736 & 0.055 & & \\
I am very good at explaining my solutions to technical problems. & 0.659 & 0.058 & & \\
I am very good at solving problems. & 0.716 & 0.059 & & \\
I am very good at coming up with new ways to solve technical problems. & 0.662 & 0.060 & & \\
I am very good at coming up with new ideas when working on projects. & 0.614 & 0.059 & & \\
\bottomrule
\addlinespace
\multicolumn{5}{l}{\textit{Note:} All factor loadings were significant ($p < .001$). } \\
\end{tabularx}
\end{table*}

Examining the relationship among the nine constructs at the bivariate correlation level (see Table \ref{tab:correlation_matrix}), all correlations were positive and statistically significant, ranging from 0.282 to 0.826, showing moderate to strong relationships between constructs. Design justice was strongly correlated to problem-solving (0.769), auditing self-efficacy (0.826), and creative expression (0.760), suggesting shared variance. Despite their high correlations, these constructs are of a distinct nature, and the model demonstrated acceptable overall fit.  

\begin{table*}[ht]
\centering
\caption{Correlation Matrix of Factors F1 through F6}
\label{tab:correlation_matrix}
\small
\begin{tabularx}{\textwidth}{l XXXXXX}
\toprule
Factor & F1 & F2 & F3 & F4 & F5 & F6 \\
\midrule
F1 - Fascination with Auditing & --- & 0.559*** & 0.282** & 0.518*** & 0.349*** & 0.369*** \\
 & & (0.078) & (0.097) & (0.083) & (0.098) & (0.095) \\
\addlinespace
F2 - Value of Learning AI/ML & & --- & 0.598*** & 0.776*** & 0.688*** & 0.688*** \\
 & & & (0.073) & (0.053) & (0.068) & (0.068) \\
\addlinespace
F3 - CS Creative Expression & & & --- & 0.676*** & 0.760*** & 0.655*** \\
 & & & & (0.066) & (0.061) & (0.071) \\
\addlinespace
F4 - Auditing Self-efficacy & & & & --- & 0.826*** & 0.745*** \\
 & & & & & (0.053) & (0.061) \\
\addlinespace
F5 - Design Justice & & & & & --- & 0.769*** \\
 & & & & & & (0.062) \\
\addlinespace
F6 - Problem-solving & & & & & & --- \\
\bottomrule
\addlinespace
\multicolumn{7}{l}{Note: $p < .01^{**}$, $p < .001^{***}$. Standard errors are in parentheses.} \\
\end{tabularx}
\end{table*}

%% file: 1_sections/5-discussion.tex
\section{Discussion}
Our paper introduces and validates an instrument for measuring students’ self-beliefs in constructing and deconstructing AI/ML systems, with initial insights into relationships between constructs. First, our analyses illustrate that it is possible to capture teens’ self-beliefs about constructing and deconstructing AI/ML systems with our survey instrument. Prior work in AI/ML instruments has largely examined how people perceive or use systems rather than how they position themselves in relation to building and evaluating them \cite{lintner2024systematic}. While recent work has started to measure self-beliefs related to justice, such work has mostly focused on adult populations, with constructs centering on how computing professionals think about justice and ethics in their work \cite{10.1145/3702652.3744208}. Our findings suggest that high school students also have structured self-beliefs in relation to AI/ML across multiple dimensions, including design justice, problem-solving, and creative expression.

These findings also provide direction for the next steps we need to take to further understand the relationships between constructs. The current correlation matrix indicates that there are strong relationships between design justice and problem-solving, auditing self-efficacy, and creative expression. These relationships suggest that beliefs about justice are not separate from self-beliefs related to construction (creative expression and problem-solving) and those related to the deconstruction of AI/ML systems (auditing self-efficacy), in other words, how students understand their ability to build and evaluate AI/ML systems. Overall, these relationships between different constructs align well with the computational empowerment framework, which sees construction and deconstruction as interconnected practices. Our research contributes to this framing by showing that justice beliefs may play a central role in both processes. However, as this analysis is correlational, the direction of these interconnections remains unclear. It is possible that an increase in design justice beliefs may predict the development of self-efficacy and problem-solving. But it is also possible that students who already feel confident in their abilities are likely to engage with concerns of justice. Future work should examine the directionality of these relationships to better understand implications for designing interventions.

Finally, the current instrument focuses on high school students. As self-beliefs are domain-specific and develop over time \cite{bandura1997self} and as the push for AI literacies expands further across different age groups \cite{mertala2022finnish, oecd2025empowering}, future work needs to examine how these constructs apply to younger learners, including middle school and elementary school students—age groups that also are interacting on a daily basis with AI/ML systems. Extending this instrument to middle or elementary contexts would also make it possible to examine whether (and how) these self-beliefs develop and shift across ages.

%% file: 1_sections/6-conclusion.tex
\section{Conclusion}

In this paper, we introduced and validated a survey instrument with constructs related to construction (creative expression and problem-solving self-beliefs) and deconstruction (auditing self-efficacy and fascination with auditing) of AI/ML systems. Our instrument was able to capture non-cognitive factors of how youth understand not only criticality but also creativity with AI/ML systems. We observed that design justice self-beliefs strongly correlated with problem-solving and creative expression, but we need further research to better understand the interconnections between these constructs. Furthermore, we suggest expanding and adapting this survey to younger groups of students to gain a better understanding of possible developmental changes over time.